# Structures and magnetic properties of iron silicide from adaptive genetic algorithm and first-principles calculations

Zejin Yang[a,b], Shunqing Wu[b,c], Xin Zhao[b], Manh Cuong Nguyen[b], Shu Yu[b,c], Tongqi Wen[b,d], Ling Tang[a,b], Fuxiang Li[b,e], Kai-Ming Ho[b] and Cai-Zhuang Wang[b*]

*a) School of Science, Zhejiang University of Technology, Hangzhou, 310023, China*

*b) Ames Laboratory, US DOE and Department of Physics and Astronomy, Iowa State University, Ames, Iowa 50011, USA.*

*c) Collaborative Innovation Center for Optoelectronic Semiconductors and Efficient Devices, Department of Physics, Xiamen University, Xiamen 361005, China*

*d) MOE Key Laboratory of Materials Physics and Chemistry under Extraordinary Conditions, School of Natural and Applied Sciences, Northwestern Polytechnical University, Xi'an 710072, China*

*e)Department of Physics, Beijing Key Laboratory of Opto-electronic Functional Materials & Micro-nano Devices, Renmin University of China, Beijing 100872 China*

**Abstract**

We performed a systematic search for low-energy structures of binary iron silicide over a wide range of compositions using the crystal structure prediction method based on adaptive genetic algorithm. 36 structures with formation energies within 50 meV/atom (11 of them are within 20 meV) above the convex hull formed by experimentally known stable structures are predicted. Magnetic properties of these low-energy structures are investigated. Some of these structures can be promising candidates for rare-earth-free permanent magnet.

## 1. Introduction

Permanent magnets (PM) are one of the essential functional materials and have been playing more and more important roles in modern technologies and energy applications[1]. Currently, most high-performance PM materials such as $Nd_2Fe_{14}B$ and $SmCo_5$ contain rare-earth (RE) elements. Due to the limited recourse and supply of the RE, it is desirable to search for replacement materials for PM without using RE. One class of the promising materials under intensive investigation is Fe-based compounds owing to the high magnetic moment, good ferromagnetic stability, low price and abundant of Fe. However, magnetocrystalline anisotropy energy (MAE) is a major concern for Fe-based compounds as pure Fe forms body-centered cubic (BCC) structure which does not exhibit any magnetic anisotropy.

*wangcz@ameslab.gov

Recently, several Fe-based RE-free compounds such as Fe-Co-X (X being B, C or N)[2-7], Fe(Co)Pt[8], and $Fe_2P$[9] have been studied intensively as potential permanent magnet materials. These studies have demonstrated that the MAE of Fe can be improved substantially by allowing with other non-magnetic elements. More effort is desirable in searching for new low-energy structures of Fe-Si compounds and characterize the magnetic properties of these structures for possible RE-free permanent magnet applications.

In this paper, we systematically explore the low-energy structures of Fe-Si compounds over a wide range of composition, especially focus on Fe-rich compounds. In addition to reproducing the experimentally synthesized structures, a number of structures with energies less than 20 meV/atom (11) and 50 meV/atom (25) with respect to the convex hull formed by experimentally know stable structures are obtained. Some of these structures have reasonably large MAE and would be potential RE-free permanent magnet materials.

## 2. Computational Methods

Low-energy structures of binary iron silicide at various compositions were searched by crystal structure prediction code using adaptive genetic algorithm (AGA).[10] In the AGA approach, auxiliary classical potentials are used to accelerate the structural search process. The parameters of the auxiliary classical potentials are adaptively adjusted by fitting to first-principles calculation results after the GA search is converged with a given interatomic potential. In each iteration of potential refitting, typically only about 20 candidate structures with lower energies from the GA search by previous auxiliary classical potential are extracted for first-principles calculations to obtain the energies and forces in order to update the classical potential parameters for next iteration of GA search. Therefore, the computer time used for the first-principles calculations is minimal. Such combined potential iteration and GA search can fully satisfy the desired calculation speed and accuracy simultaneously in the structure search using genetic and related algorithms.

In the present study, crystal unit cells contain up to 8 formula units (f.u.) are used in the AGA search. The size of the structure pool in the AGA search is 128 and the search with a given interatomic potential is satisfied to be converged if the lowest energy 128 structures remain unchanged in 500 consecutive GA generations. Then 16 lowest-energy structures at the end of each GA search with a given set of classical potential parameters are selected for first-principle calculations to update the potential parameters for next iteration of GA search, and so on. The interatomic potentials based on the Embedded Atom Method (EAM) formalism are adopted in the AGA searches[11]. The potential fitting is done by the force-matching method with stochastic simulated annealing algorithm adopted in the potfit code[12,13].

The first-principles calculations are carried out using the projector-augmented wave method[14] with a plane wave basis set, as implemented in the VASP[15,16] code. The spin-polarized generalized gradient approximation and the Perdew-Burke-Ernzerhof[17] functional are used to model the exchange-correlation energy. Energy cutoff for the plan-wave basis is 350 eV. The self-consistent convergence criterion for the energy is $1.0\times10^{-7}$ eV. K-point sampling[18] of $2\pi\times0.013$ Å$^{-1}$ in the Brillouin-zone is used and the force on each atom is converged to less than 0.01 eV/Å during the atomic relaxation. After the crystal structure search and optimization, low-energy structures are selected for magnetic property (including MAE) calculations. Spin-orbit coupling is included in the MAE calculation and the symmetry operation is switched off (ISYM =-1) during the MAE calculations.

## 3. Results and discussion

### 3.1 Low-energy structures

The formation energies of the low-energy structures of binary $Fe_xSi_y$ compounds with different compositions at zero pressure and temperature obtained from our AGA search and first-principles total energy calculations are shown in Fig. **1**. The convex hull is constructed using the diamond Si and BCC Fe crystal structures as well as several experimentally determined structures of binary Fe-Si compounds, *i.e.*, ($FeSi_2$, FeSi, and $Fe_3Si$). The formation energies are all negative suggesting that these structures are stable with respect to the decomposition into elemental Si and Fe crystals. However, all the formation energies of the structures from the AGA search are above the convex hull indicating that these structures would be metastable structures.

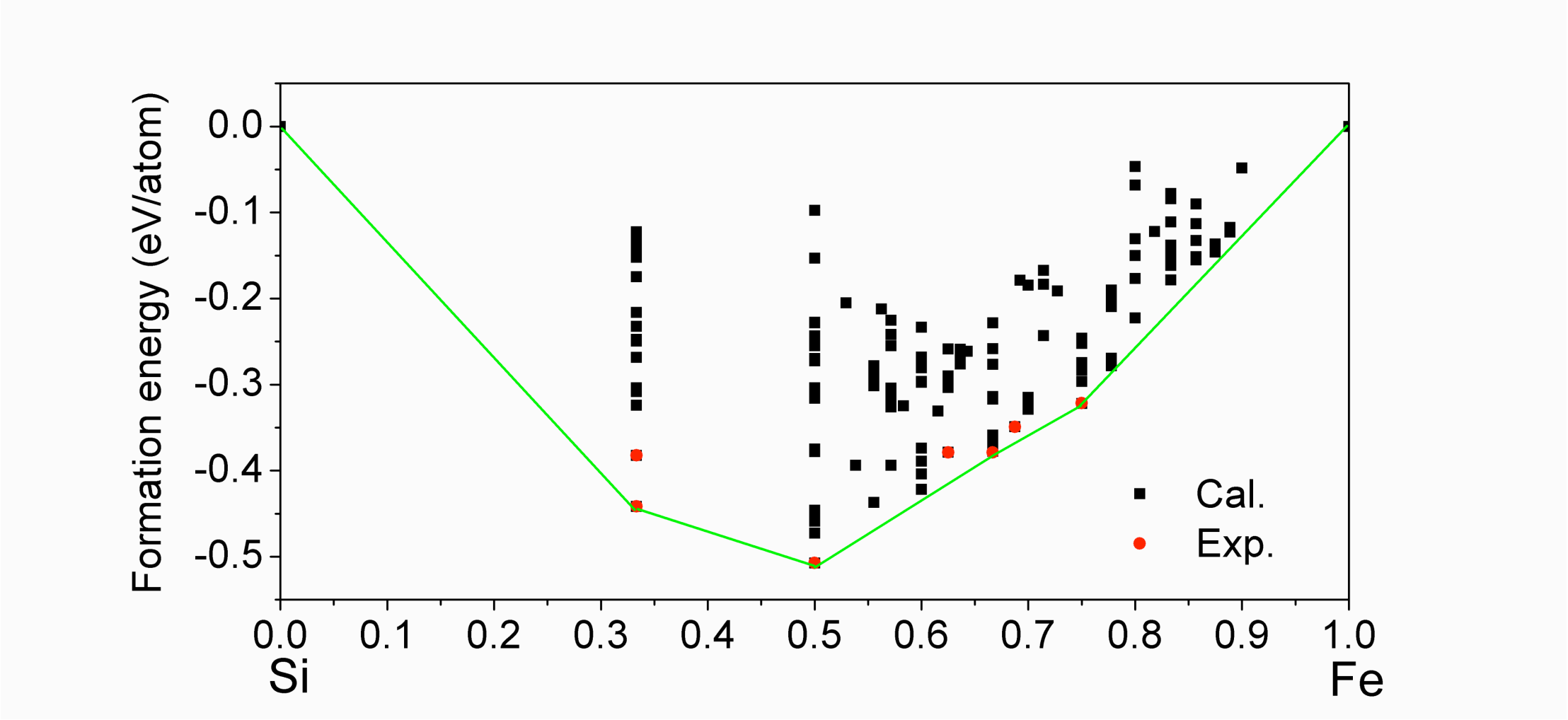

**Figure 1**. The formation energies of $Fe_xSi_y$ obtained from our AGA search. The formation energy is calculated as $[E(Fe_xSi_y)-xE(Fe)-yE(Si)]/(x+y)$, *E* means energy. The convex hull is constructed using the energies of the diamond Si , BCC Fe as well as experimentally determined structures of $FeSi_2$, FeSi, and $Fe_3Si$. The black squares are the results from the AGA structures and the red open circles are from the experimentally observed structures. The lowest-energy structures at $FeSi_2$, FeSi, and $Fe_3Si$ are also reproduced by our AGA search.

Recently, materials synthesis under far-from-equilibrium conditions has become an important approach for materials design and discovery. Various far-from-equilibrium synthesis methods have been developed and used to obtain metastable structures which could possess desirable properties and functionalities. In the literature, many metastable structures with formation energy 100 meV/atom or higher above the corresponding convex hull have been observed by experiments under non-equilibrium synthesis conditions. In binary iron silicide system, 7 compounds have been experimentally synthesized so far, including three metastable phases, *i.e.*, $FeSi_2$ ($P4/mmm$), $Fe_5Si_3$ ($P6_3/mcm$), $Fe_{11}Si_5$ ($Pm\bar{3}m$), with energy about 60, 30, and 10 meV/atom respectively with respective to the corresponding ground-state structures. Our calculations show that $Fe_5Si_3$ ($P6_3/mcm$) exists a large easy plane magnetic anisotropy with a MAE value of 2.21 MJ/m$^3$ ($E_{100}$=$E_{010}$<$E_{001}$), whereas $Fe_2Si$ ($P\bar{3}m1$) shows a relatively smaller MAE, with a value of 0.51 MJ/m$^3$ ($E_{100}$=$E_{010}$<$E_{001}$). Other experimentally observed iron silicides, including $FeSi_2$ ($Cmce$, $P4/mmm$), FeSi ($P2_13$), $Fe_3Si$ ($Fm\bar{3}m$), $Fe_{11}Si_5$ ($Pm\bar{3}m$), have zero MAE.

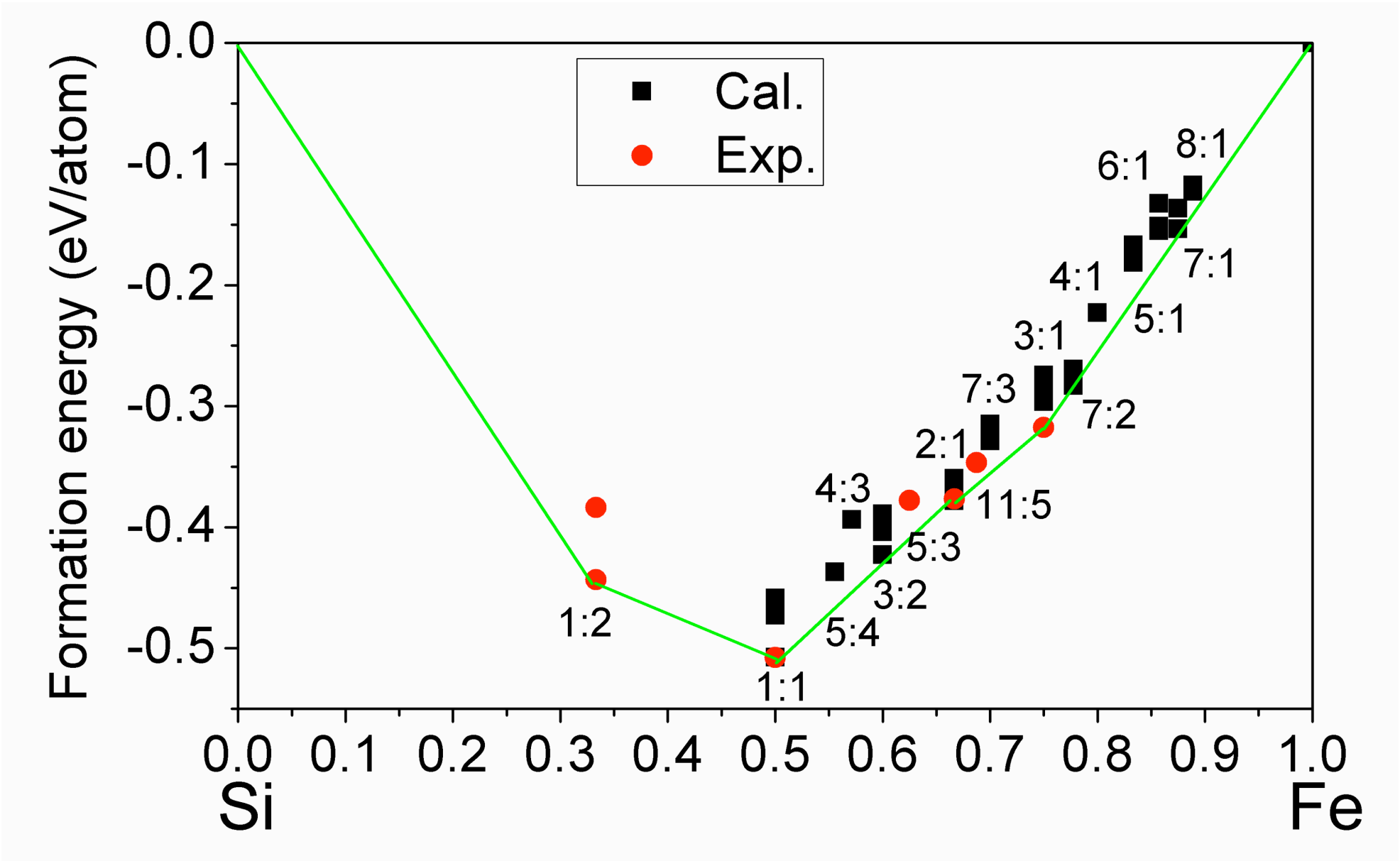

**Figure 2** The formation energies of the $Fe_xSi_y$ structures with the formation energy less than 50 meV/atom above the convex hull obtained from our AGA search.

Since the structures with formation energy as high as 60 meV/atom above the convex hull have been observed experimentally, some of the low-energy metastable structures obtained from our AGA search would be synthesized by experiment. In Fig.

2, we show all the metastable structures with the formation energy less than 50 meV/atom above the convex hull obtained from our AGA search. Among these 36 low-energy metastable structures, 11 of them have formation energies less than 20 meV/atom above the convex hull. There is a high chance that some of these 36 low-energy metastable structures can be made by experiment through the non-equilibrium synthesis techniques. In the following sections, our discussion on the magnetic properties of $Fe_xSi_y$ compounds will base only on these 36 low-energy structures.

### 3.2 Magnetic properties

We have calculated the magnetic properties and MAE for the 36 lowest-energy (less than 50 meV above the convex hull) metastable structures. The results are listed in Table **1**.

**Table 1**. Structural and magnetic properties of the 11 low-energy structure (within 20 meV/atom from the convex hull) of $Fe_xSi_y$. The magnetic moments are in the unit of $\mu_B$. The MAE (μeV/atom) are calculated along three directions as indicated. n.f.u. means the number of formula units used in the AGA search to obtain the structures.

| composition | Symmetry | MAE | 100 | 010 | 001 | Mag. Momenet per atom | Mag. Moment per Fe | n.f.u. |
|---|---|---|---|---|---|---|---|---|
| FeSi | $Pm\bar{3}m$ | none | 0 | 0 | 0 | 0 | 0 | 4 |
| $Fe_3Si_2$_A | $C2/m$ | small | 0 | 3.7 | 0.5 | | | 4 |
| $Fe_2Si$ | $Cmcm$ | small | 13.69 | 5.78 | 0 | 0.70 | 1.05 | 4 |
| | $P4/nmm$ | none | 0 | 0 | 0 | 0.71 | 1.07 | 4 |
| | $C2/m$ | small | 0 | 4.03 | 1.53 | 0.69 | 1.04 | 4 |
| | $P2/m$ | small | 0 | 3.50 | 0.18 | 0.67 | 1.01 | 4 |
| | $P4/mmm$ | easy | 3.39 | 3.39 | 0 | 0.70 | 1.05 | 4 |
| | | axis | | | | | | |
| $Fe_7Si_2$ | *Pm* | small | 0 | 12.36 | 3.97 | 1.41 | 1.82 | 4 |
| | *Cmmm* | large | 30.30 | 25.63 | 0 | 1.38 | 1.78 | 2 |
| $Fe_7Si$ | *P*1 | small | 13.33 | 24.24 | 0 | 1.85 | 2.12 | 4 |
| $Fe_8Si$ | $P\bar{1}$ | small | 9.15 | 0 | 10.47 | 1.93 | 2.17 | 4 |

Table 2 Structural and magnetic properties of the 25 structures of $Fe_xSi_y$ with the energies between 20 to 50 meV/atom above the convex hull. Same definitions are used as those in **Table 1**.

| Composition | Symmetry | MAE | 100 | 010 | 001 | Mag. Moment per atom | Mag. Moment per Fe | nf.u. |
|---|---|---|---|---|---|---|---|---|
| FeSi | $P\bar{6}m2$ | small | 0 | 14.75 | 26.75 | 0.37 | 0.75 | 4 |
| | *C*2/*m* | none | 0 | 0 | 0 | 0 | 0 | 4 |
| $Fe_5Si_4$ | *P*2/*m* | small | 11.58 | 18.17 | 0 | 0.25 | 0.45 | 4 |
| $Fe_4Si_3$ | *I*4/*mmm* | large | 49.61 | 0 | 41.17 | 0.37 | 0.65 | 4 |
| $Fe_3Si_2$ | *P4/m* | easy axis | 20.44 | 0 | 20.44 | 0.42 | 0.71 | 4 |
| | $P2_1/m$ | nearly perfect | 27.13 | 27.42 | 0 | 0.47 | 0.79 | 4 |
| $Fe_3Si_2$_B | *C*2/*m* | large | 58.70 | 47.21 | 0 | 0.58 | 0.98 | 4 |
| $Fe_7Si_3$ | $P2_1/m$ | small | 0 | 24.29 | 11.93 | 0.95 | 1.36 | 4 |
| | *Pm* | small | 0 | 32.78 | 1.42 | 0.88 | 1.26 | 4 |
| | $P\bar{1}$ | small | 12.47 | 32.13 | 0 | 0.88 | 1.26 | 4 |
| $Fe_3Si$ | *Fmmm* | small | 0.88 | 0.09 | 0 | 1.276 | 1.70 | 4 |
| | *Cmcm* | small | 25.94 | 6.02 | 0 | 1.20 | 1.60 | 4 |
| | *C*2/*m* | small | 0 | 12.37 | 4.23 | 1.44 | 1.93 | 4 |
| | *C2* | nearly perfect | 10.33 | 9.98 | 0 | 1.06 | 1.41 | 8 |
| $Fe_4Si$ | *C*2/*m* | small | 0 | 11.35 | 21.32 | 1.94 | 2.43 | 4 |
| $Fe_5Si$ | *P*1 | small | 1.10 | 17.86 | 0 | 1.76 | 2.11 | 4 |
| | $P2_1/m$ | small | 62.64 | 12.59 | 0 | 1.76 | 2.11 | 4 |
| | *C*2/*m* | small | 11.88 | 0 | 49.37 | 1.74 | 2.09 | 4 |
| *$Fe_6Si$* | *Cmmm* | easy axis | 27.14 | 27.14 | 0 | 1.57 | 1.83 | 4 |
| | $P\bar{1}$ | small | 16.00 | 10.11 | 0 | 1.79 | 2.09 | 4 |
| | *I*4*mm* | easy axis | 6.04 | 6.04 | 0 | 1.64 | 1.91 | 4 |
| | *P*1 | small | 14.47 | 0 | 21.44 | 1.83 | 2.14 | 4 |
| | *Pm* | small | 23.28 | 0 | 11.55 | 1.77 | 2.07 | 4 |
| $Fe_7Si$ | $P\bar{1}$ | small | 8.80 | 12.18 | 0 | 1.88 | 2.15 | 4 |
| $Fe_8Si$ | *P*1 | small | 0 | 4.28 | 2.68 | 1.90 | 2.14 | 4 |

According to the values of MAE along different crystalline directions, we classify these structures into three groups. Group I refers to those structures with uniaxial anisotropy, i.e., only one direction has lower energy and energies along other directions are the same and higher than that of the easy axis. Group II includes those structures have large energy difference along the different directions but the energies along the two higher energy directions are not the same. We refer such situation as triaxial anisotropy and the MAE is the energy difference between the two lower-energy directions. Group III are those structures with very small or even zero MAE. The structures of the group I and II are shown in Fig. **3** and **4** respectively.

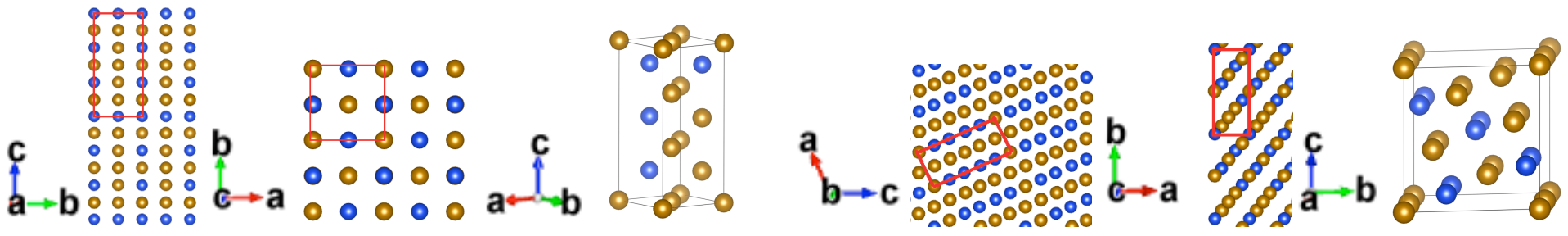

**(a)** $Fe_2Si$_ *P4/mmm* 3.39 **(b)** $Fe_3Si_2$_ *P4/m* 20.4

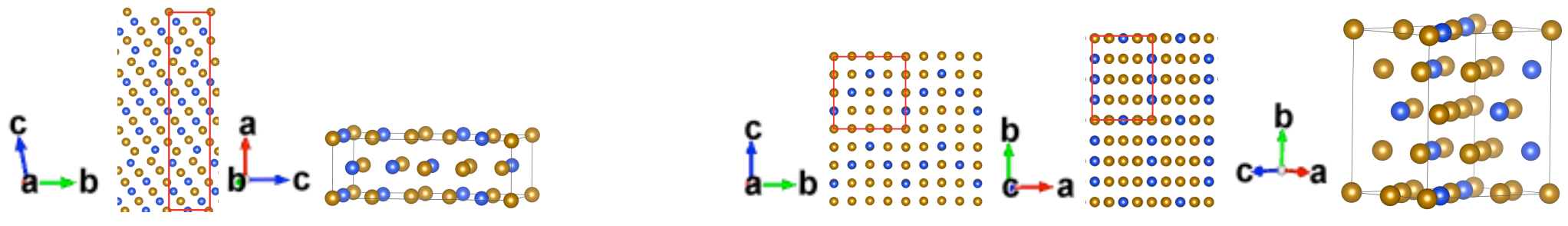

**(c)** $Fe_3Si_2$_ *P*2$_1$*/m* 27.13 **(d)** $Fe_3Si$_*C2* ~10.0

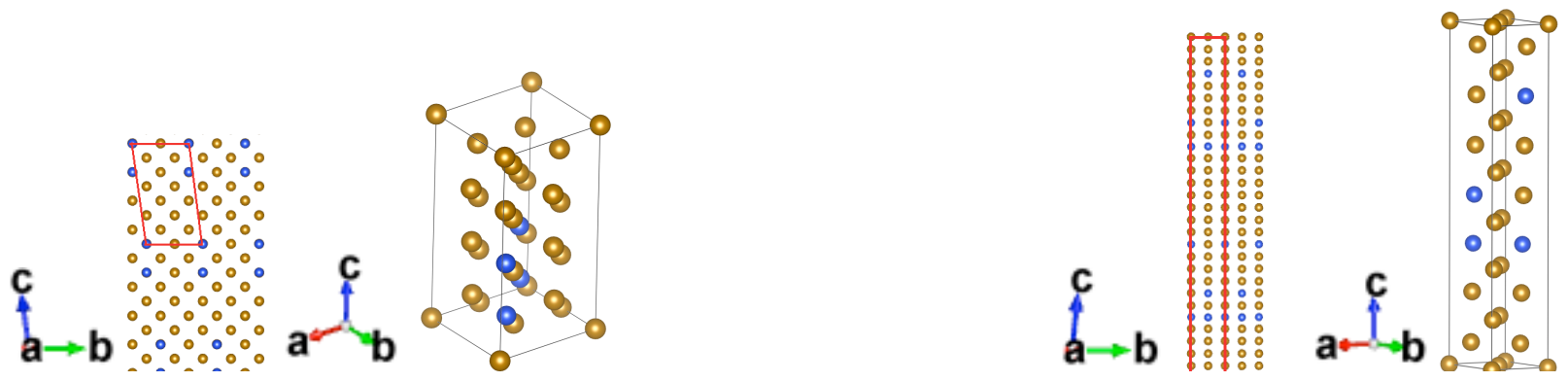

**Figure 3** Structures with uniaxial magnetic anisotropy, digital in the picture means MAE with unit of μeV, Si Fe.

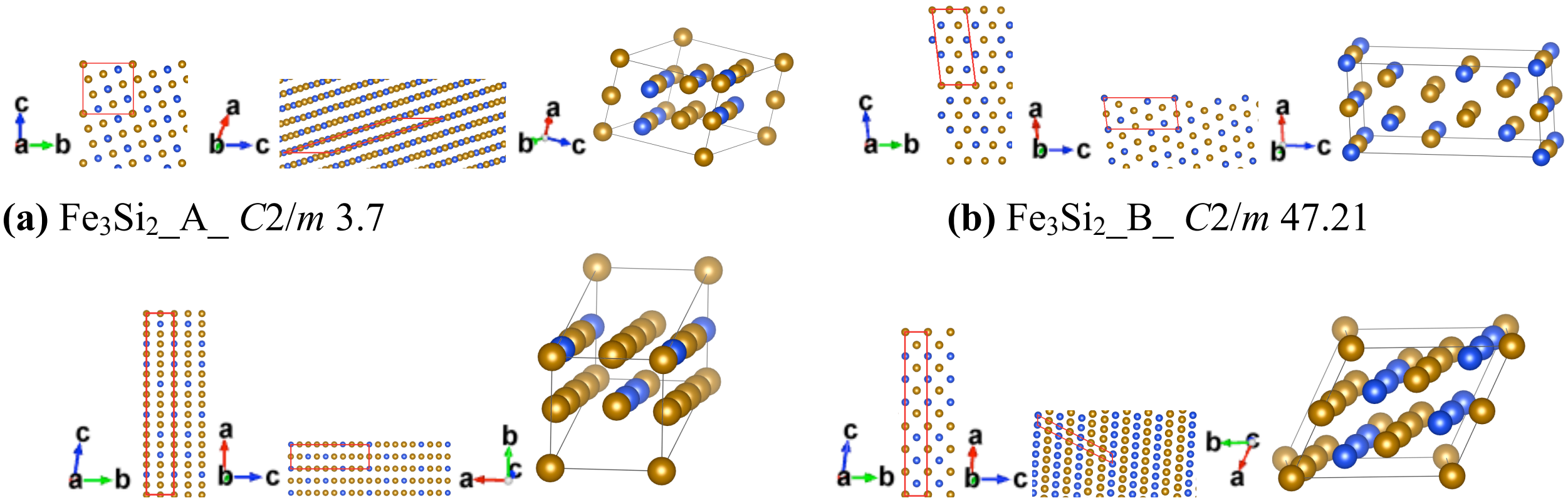

**(a)** $Fe_3Si_2$_A_ *C*2/*m* 3.7 **(b)** $Fe_3Si_2$_B_ *C*2/*m* 47.21

**(c)** $Fe_7Si_2$_*Cmmm* 25.63 **(d)** $Fe_4Si_3$_*I*4/*mmm* 41.17

**Figure 4** Structures with triaxial magnetic anisotropy, digital in the picture means MAE with unit of μeV, Si Fe.

### 3.2.1 Structures with relatively large MAE

The $Fe_3Si_2$ (*P*4/*m*) is a tetragonal crystal with structural parameters $a$=$b$=6.47 Å, $c$=2.71 Å, $\alpha$=$\beta$=$\gamma$=90°, and with two formula units per primitive cell, as is shown in Fig. **3** (b). The $Fe_6Si$ (*Cmmm*) shown in Fig. **3** (c) has lattice parameters $a$=$b$=5.58 Å, $c$=10.14 Å, $\alpha$=$\beta$=97.91°, $\gamma$=90.0°, and with two formula units per primitive cell. Both structures have energies larger than 20 meV/atom with respect to the convex hull. These two structures exhibit uniaxial magnetic anisotropy with MAE of 0.31 and 0.39 $MJ/m^3$, respectively. The average atomic magnetic moment in unit cell of the $Fe_3Si_2$ is 0.4281 $\mu_B$ although two Fe atoms in the unit cell have very large moment of 2.758 $\mu_B$. By contrast, $Fe_6Si$ structure has more evenly distributed moment of 1.57 $\mu_B$ per atom and 1.83 $\mu_B$ per Fe atom in the unit cell.

Four other structures ($Fe_2Si$_*P4/mmm*, $Fe_3Si_2$_ $P2_1/m$, $Fe_6Si$_ $I4mm$, and $Fe_3Si$_*C*2) are also shown in **Fig. 3**. Although the $Fe_2Si$_*P4/mmm* exhibit uniaxial anisotropy, its MAE is very small, 3.39 μeV/atom. The average magnetic moment is 0.7 per atom and 1.05 $\mu_B$ per Fe atom. Nevertheless, the formation energy of the $Fe_2Si$_ *P4/mmm* structure is very close to the convex hull, (only about 19 meV/atom above the convex hull) although it is not the lowest-energy one at the composition obtained from our AGA search (see Table **1**). $Fe_3Si_2$_$P2_1/m$ has moderate MAE of about 27 μeV/atom (0.41 $MJ/m^3$) probably due to its large distortion from the BCC lattice. However, the magnetic moment of this structure is relatively small (0.47 per atom and 0.97 $\mu_B$ per Fe) due to relatively high concentration of Si. The MAE of $Fe_6Si$_$I4mm$ structure is very small (see Table **2**) and could be ignored due because the lattice is still nearly BCC. However, the average magnetic moment is high, 1.64 per atom or 1.91 $\mu_B$ per Fe atom due to its Fe-rich composition. Finally, the $Fe_3Si$_*C*2 has MAE of about 10.0 μeV/atom (0.15$MJ/m^3$) and average magnetic moment of 1.06 per atom or 1.41 $\mu_B$ per Fe atom. More details of the structural properties along with other low-energy structures can also be found in the supporting materials.

In Fig. **3**, $Fe_3Si_2$ (*P*4/*m*) shows perfect MAE along *a* and *c* axes (20.4 μeV/atom) with same projections along *a* and *b* axes. $Fe_6Si$ (*Cmmm*) has perfect MAE along *a* and *b* axes, with identical values of 27.14 μeV/atom, in comparison with that of lower energy direction *c* axis, consisting with the fact of same projections along *a* and *b* axes, namely, the unit cell has a $C_{2V}$ mirror symmetry with respect to the plane formed by *c* axis. $Fe_2Si$_*P4/mmm* shows perfect easy axis but with nearly zero value, 3.39 μeV along 100 and 010 axes relative to that of 001 axis. The energy difference between the $Fe_2Si$_*P4/mmm* with the stable state is below 20 meV/atom. Only this phase is within the energy range of 0~20 meV/atom among all the studied 36 structures, other structures with perfect or nearly perfect easy-axis are within 20~50 meV/atom from their individual stable states. Its atomic distribution features are same along *a* and *b* axes, agreeing with the fact of the identical energies along *a* and *b* axes. $Fe_3Si_2$_ $P2_1/m$ shows nearly perfect easy axis, with respective values of 27.13 and 27.42 μeV along 100 and 010 axes relative to that of 001 direction. Probably the distorted BCC building block is the origin of the imperfect MAE of the crystal. $Fe_6Si$_$I4mm$ shows nearly perfect easy axis but with nearly zero value, 6.0471 μeV along 100 and 6.0367 μeV along 010 axes relative to that of 001 axis. The unit cell shows $C_{2v}$ planar symmetry with respect to the plane formed by *c* axis, namely, projections towards *a* and *b* are same, the nearly identical MAE could be reflected by their totally symmetric geometric positions in regardless of the elemental species. $Fe_3Si$_*C*2 shows a slightly large difference between 100 and 010 axes, with respective values of 10.33 and 9.98 μeV relative to that of 001 axis, similar with the case of $Fe_6Si$_$I4mm$.

We also found that some structures with relatively large MAE but the anisotropy is not uniaxial. Among the selected structures with energies below 20 or within 20-50 meV/atom to their respective stable states, only three structures belong to these categories. These structures are $Fe_3Si_2$_B_$C2/m$, $Fe_7Si_2$_$Cmmm$, and $Fe_4Si_3$_$I4/mmm$, respectively, as shown in Fig. **4** (in which $Fe_3Si_2$_A_$C2/m$ is only reference for $Fe_3Si_2$_B_*C*2/*m*). The structural parameters of the $Fe_3Si_2$_B_ *C*2/*m* are parameters 4.78 (*a*=*b*) Å, 10.05 (*c*) Å, 94.83° (*α*=*β*), 109.59° (*γ*), respectively, and this structure has large MAE along 100 and 010 axes relative to 001 axis, with values of 58.7 and 47.21 μeV/atom, respectively. $Fe_7Si_2$_*Cmmm* structure with structural parameters of 3.94 (*a*=*b*) Å, 12.94 (*c*) Å, 98.76° (*α*), 90.0° (*β*=*γ*), respectively shows MAE 30.3 and 25.63 μeV/atom along 100 and 010 axes, respectively. $Fe_4Si_3$_ *I*4/*mmm* is a structure with lattice parameters 6.14 (*a*=*c*) Å, 10.75 (*b*) Å, 73.38° (*α*), 53.13° (*β*), 93.28° (*γ*) and shows MAE of 49.61 and 41.17 μeV/atom along 100 and 001 axes, respectively. Their magnetic moments oth these structures can be found in Table **2**.

We note that 25 structures with formation energy within 50 meV/atom above the convex hull exhibit uniaxial magnetic anisotropy but the MAE are small. These structures would be good candidates for soft ferromagnetic materials. These structures include $P\bar{6}m2$, *C*2/*m*, *P*2₁/*c* of FeSi, *P*2/*m* of $Fe_5Si_4$, *I*4/*mmm* of $Fe_4Si_3$, $P2_1/m$ of $Fe_3Si_2$, *P*2/*m*, $P2_1/m$, $Pm$, $P\bar{1}$ of $Fe_7Si_3$, $Fmmm$, $Cmcm$, *C*2/*m*, *C*2 of $Fe_3Si$, *C*2/*m* of $Fe_4Si$, $P1$, $P2_1/m$, *C*2/*m* of $Fe_5Si$, $Cmmm$, $P\bar{1}$, $I4mm$, $P1$, $Pm$ of $Fe_6Si$, $P\bar{1}$ of $Fe_7Si$, $P1$ of $Fe_8Si$, respectively. Detailed structural and magnetic properties of these structures are shown in Tables **1** and **2** and supporting information.

**3.3.2 Effect of Si on the magnetic moment of Fe atoms**

Analysis based on the results obtained from our calculations show that alloying with Si has profound effects on the magnetic moments of Fe atoms. In general, the overall magnetic moments of the $Fe_xSi_y$ compounds are smaller than that of the BCC Fe crystals and the average moment decease with the increase of the Si composition.

However, some individual Fe atoms can have magnetic moment higher than their values in BCC crystal. We find that the magnetic moment of individual Fe atom is strongly dependent on its chemical as well as structural environments. In general, the magnetic moment of the central Fe atom increases as the percentage of Fe atom in its first neighbor shell increases, as shown in Fig. **5**. Some Fe atoms can even achieve the moment of 2.5-3.0 $\mu_B$. A close inspection of Fig. **5** shows that although the magnetic moment of the central Fe atom increases with the number of Fe atoms in its first shell, the variation of the magnetic moment at given Fe coordination numbers can be quite spread. For example, the magnetic moments at the first-shell Fe ratio of 0.5 range from about 0.5 to 1.5 $\mu_B$. These results suggest that the magnetic moment of the central atom can also be dependent on the way how the Fe (or Si) atoms are distributed in the first shell. As an illustration, we show in Fig. **6** that different chemical arrangement around the central Fe atom can give very different magnetic moment from about 0.5 to 1.5 $\mu_B$ although the ratios of the Fe atoms in the first shell are the same in the three cases. It seems that more asymmetric distribution of the Fe (or Si) atoms will give higher magnetic moment to the central Fe atom.

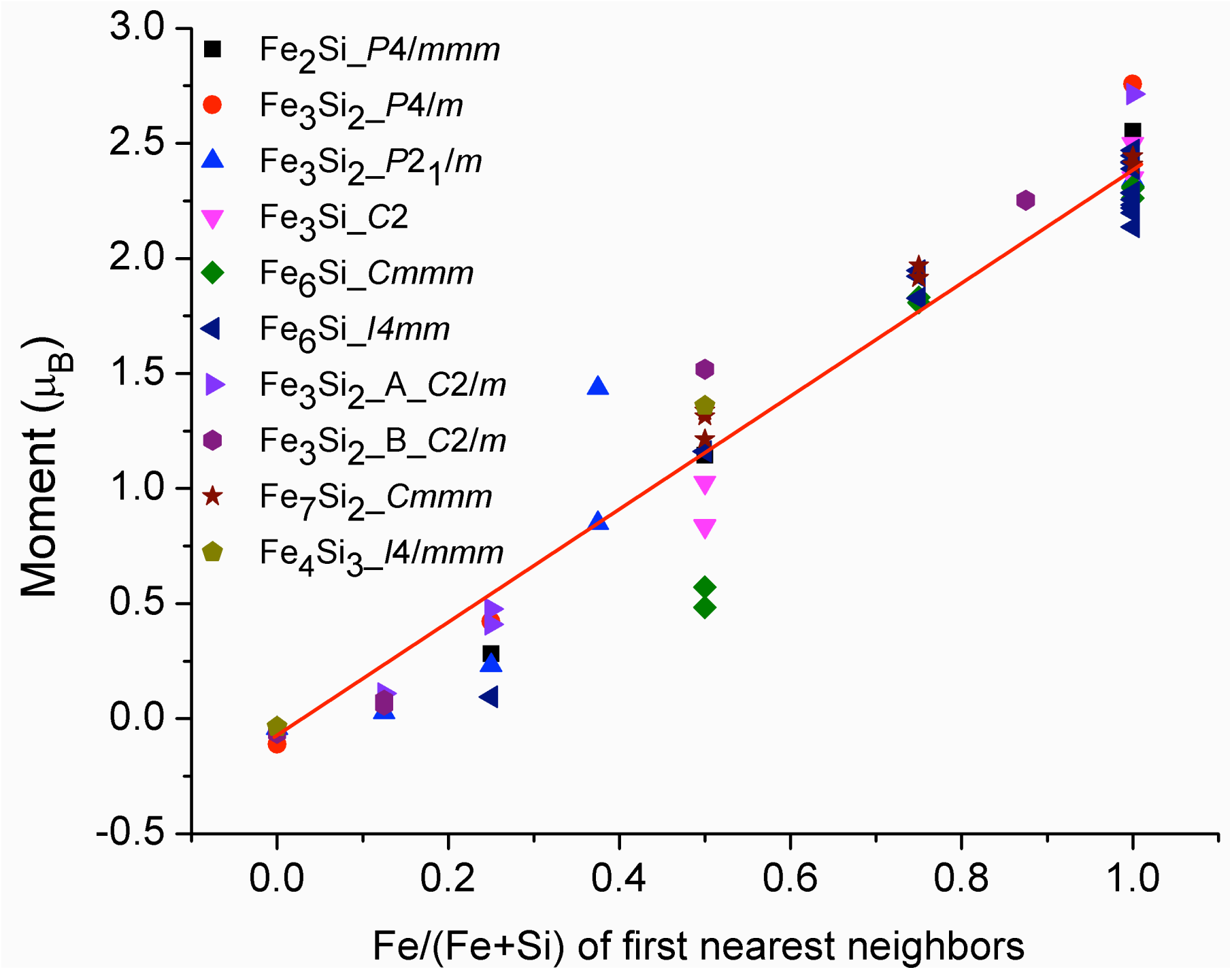

**Figure 5** Magnetic moment of the central Fe atom in the $Fe_xSi_y$ compounds as the function of the percentage of Fe atom in its first neighbor shell.

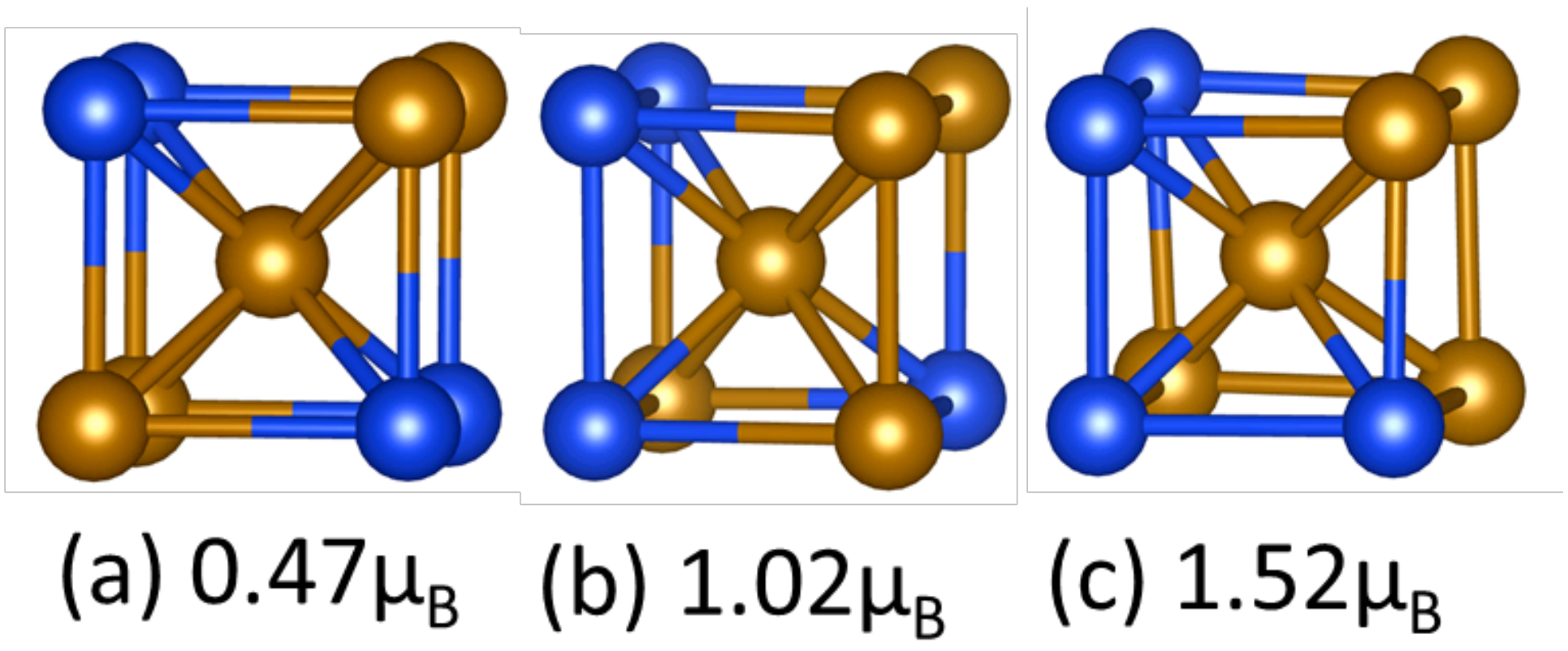

**Figure 6** Dependence of the magnetic moment of the central Fe atom on the chemical destruction of the first shell atoms. Note that the ratio of Fe:Si in the first shell is the same for three structures, Si Fe.

The magnetic moment of the Fe atom is also dependent on the distance from its first neighbor atoms, as shown in Fig. **7**. The longer the average distance, the higher the magnetic moment.

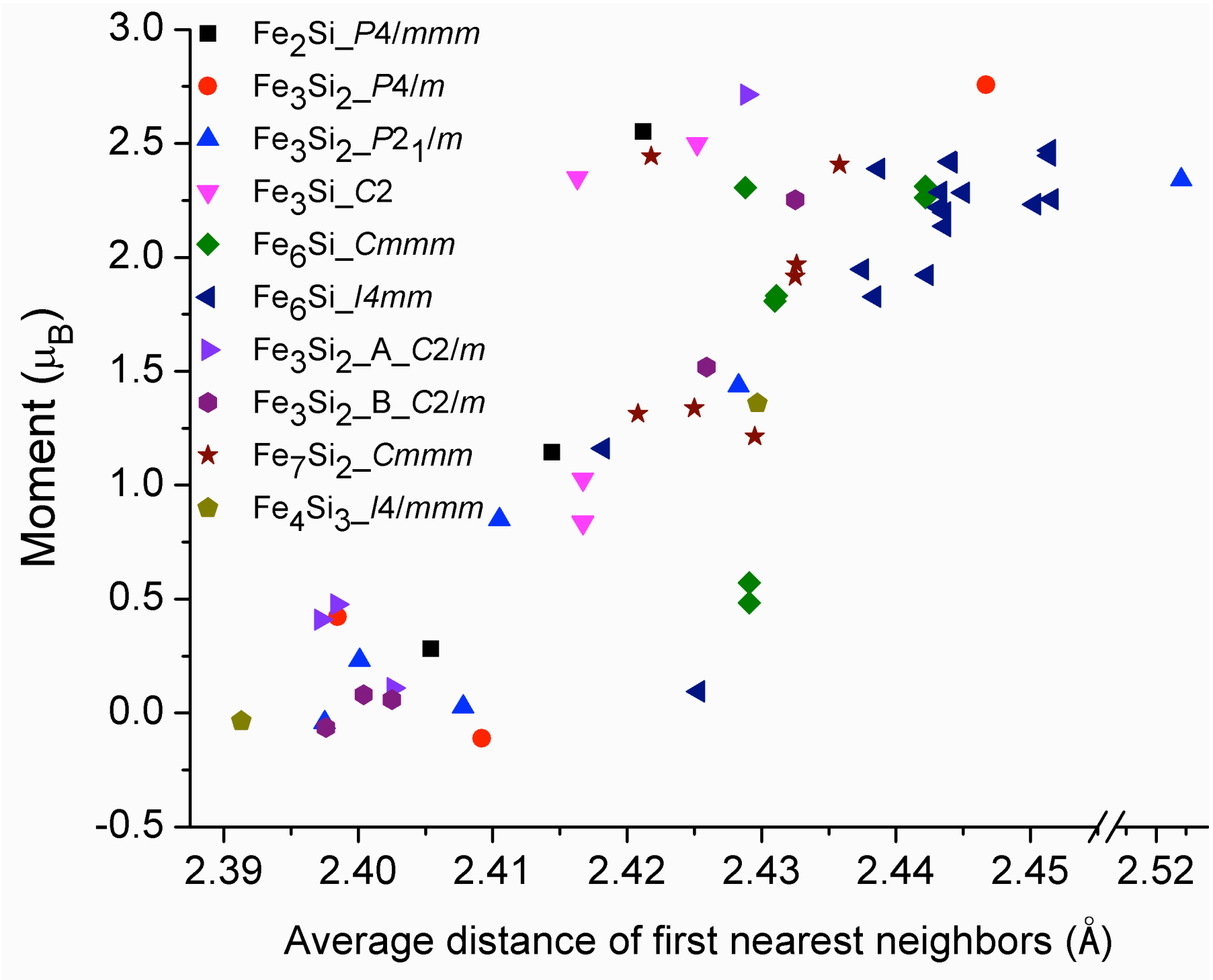

**Figure 7** Magnetic moment of the central Fe atom in the $Fe_xSi_y$ compounds as the function of the average bond lengths between the central atom and its first neighbor shell atoms.

It should be noted that even for a given crystal structure, the magnetic moment on each Fe atom can be very different because the chemical environment can be different for each individual Fe atoms in the same crystal structure. An example is shown in Fig. **8**.

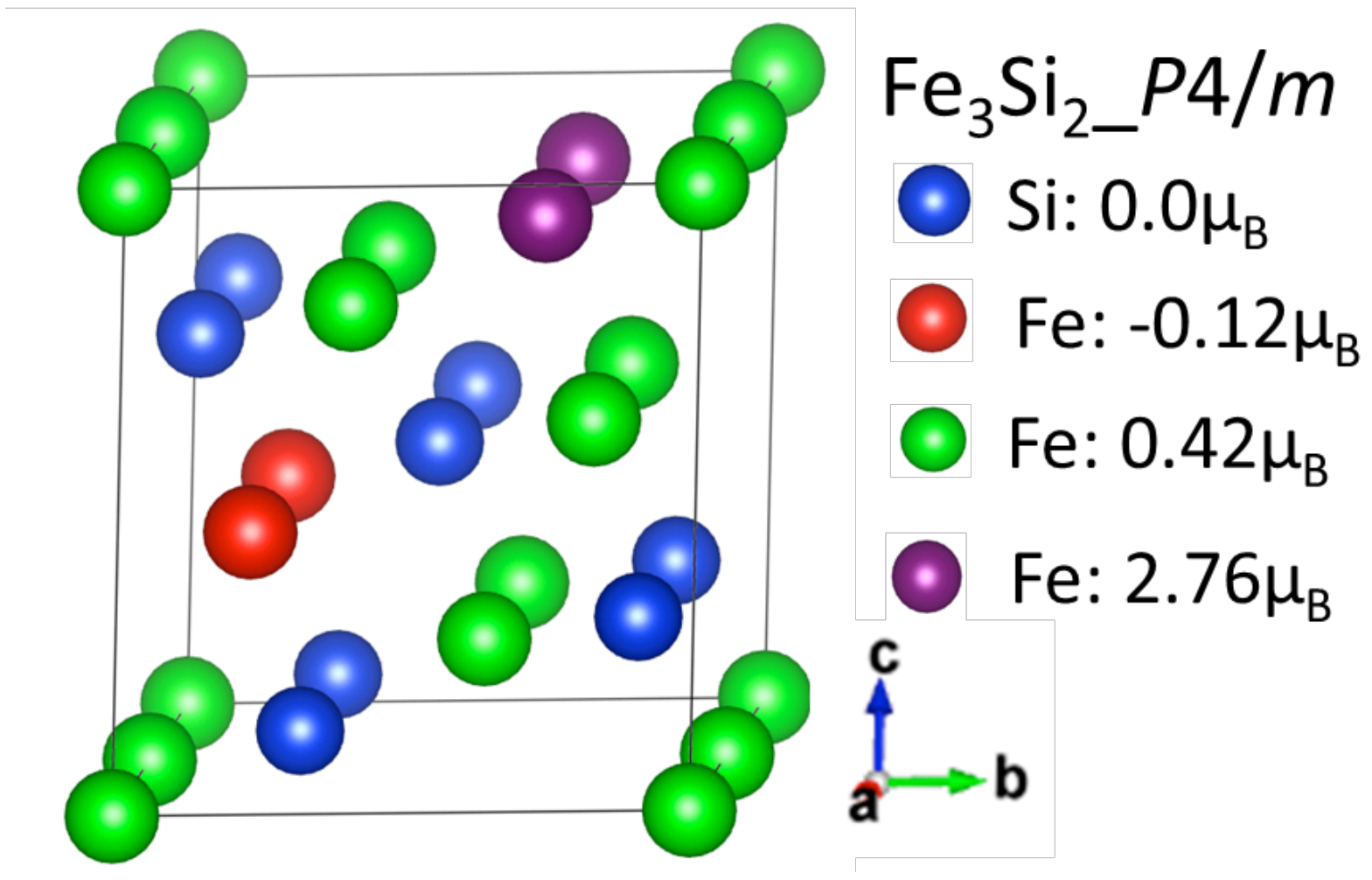

**Figure 8** Structure and moment ($\mu_B$) of typical phases $Fe_6Si$_*Cmmm*

The lattice of the $Fe_3Si_2$_*P*4/*m* structure shown in Fig. **8** is BCC. There three types of Fe atoms (colored red, green, and purple respectively) in this structure according to the distribution of Si atoms (blue) around them. The red Fe atoms are all surround by Si atom in its first shell, therefore they have very small magnetic moment of -0.12 $\mu_B$ (even the direction of the moment is different from the other two types). The green Fe atoms have more Si atoms than Fe atoms in their first neighbor shell and their magnetic moment is also small (0.12 $\mu_B$). By contrast, because there are no Si atoms in their first neighbor shell, the purple Fe atoms exhibit very large magnetic moment of 2.76 $\mu_B$, which is even larger than that of Fe in BCC pure-Fe crystal.

## 4. Conclusion

Using the adaptive genetic algorithm and first principles calculations, we systematically explored the magnetic properties of the low-energy $Fe_xSi_y$ compounds over a wide range of chemical compositions. 36 new structures with energies less than 20 meV/atom (11) and 50 meV/atom (25) with respect to the convex hull formed by experimentally know stable structures are determined. The deeply discussed structures of the 36 predicted metastable structures have body-centered cubic (BCC) building blocks with some degree of distortion from the BCC lattice. The presence of Si in the $Fe_xSi_y$ compounds break the chemical symmetry of the BCC lattice and lead to some moderate magnetic anisotropy in some of the binary compounds. We also show that the magnetic moment of the Fe atoms in $Fe_xSi_y$ compounds is also sensitive to the chemical and structural environment of the central Fe atoms. The information obtained from such a systematic study will be very useful for guiding experiment in design and synthesis of Fe-Si-based materials for magnetic applications.

**Acknowledgements**

This work was supported by the National Science Foundation, Division of Materials Research under Award DMREF: SusChEM 1436386. The development of adaptive genetic algorithm method was supported by the US Department of Energy, Basic Energy Sciences, Division of Materials Science and Engineering, under Contract No. DE-AC02-07CH11358, including a grant of computer time at the National Energy Research Scientific Computing Center in Berkeley, CA. Zejin Yang acknowledges the financial support from China Scholarship Council (File No. 201508330131) and the hospitality of Ames lab and Iowa State University. Zejin Yang acknowledges the Natural Science Foundation of Zhejiang Province, China (Grant No: LY18E010007).